\documentclass[]{aastex701}
\usepackage{graphicx}

\begin{document}

\title{Deep imaging of the very isolated dwarf galaxy NGC6789}

\author[orcid=0000-0001-8647-2874]{Ignacio Trujillo}
\affiliation{Instituto de Astrof\'{i}sica de Canarias, V\'{i}a L\'{a}ctea S/N, E-38205 La Laguna, Spain}
\affiliation{Departamento de Astrof\'{i}sica, Universidad de La Laguna, E-38206 La Laguna, Spain}
\email[show]{trujillo@iac.es}  

\author[orcid=0009-0001-7407-2491]{Sergio Guerra Arencibia} 
\affiliation{Instituto de Astrof\'{i}sica de Canarias, V\'{i}a L\'{a}ctea S/N, E-38205 La Laguna, Spain}
\affiliation{Departamento de Astrof\'{i}sica, Universidad de La Laguna, E-38206 La Laguna, Spain}
\email{sguerra@iac.es}

\author[orcid=0009-0003-6502-7714]{Ignacio Ruiz Cejudo}
\affiliation{Instituto de Astrof\'{i}sica de Canarias, V\'{i}a L\'{a}ctea S/N, E-38205 La Laguna, Spain}
\affiliation{Departamento de Astrof\'{i}sica, Universidad de La Laguna, E-38206 La Laguna, Spain}
\email{ignaciomrrd.es@gmail.com}

\author[orcid=0000-0001-7847-0393]{Mireia Montes}
\affiliation{Institute of Space Sciences (ICE, CSIC), Campus UAB, Carrer de Can Magrans, s/n, 08193 Barcelona, Spain}
\email{mmontes@ice.csic.es}

\author[0000-0002-8134-2592]{Miguel R. Alarcon}
\affiliation{Instituto de Astrof\'{i}sica de Canarias, V\'{i}a L\'{a}ctea S/N, E-38205 La Laguna, Spain}
\affiliation{Departamento de Astrof\'{i}sica, Universidad de La Laguna, E-38206 La Laguna, Spain}
\affiliation{Light Bridges S.L., Observatorio del Teide, Carretera del Observatorio s/n, E-38500 Guimar, Tenerife, Canarias, Spain}
\email{miguelrguez.alarcon@gmail.com}

\author[0000-0002-2394-0711]{Miquel Serra-Ricart}
\affiliation{Instituto de Astrof\'{i}sica de Canarias, V\'{i}a L\'{a}ctea S/N, E-38205 La Laguna, Spain}
\affiliation{Departamento de Astrof\'{i}sica, Universidad de La Laguna, E-38206 La Laguna, Spain}
\affiliation{Light Bridges S.L., Observatorio del Teide, Carretera del Observatorio s/n, E-38500 Guimar, Tenerife, Canarias, Spain}
\email{miquel@lightbridges.es}

%\collaboration{all}{The Terra Mater collaboration}

\begin{abstract}

We present deep optical imaging of the extremely isolated dwarf galaxy NGC 6789, obtained with the new 2-meter Two-meter Twin Telescope (TTT3) at Teide Observatory. Despite its location in the Local Void, NGC 6789 exhibits surprising recent central star formation equivalent to approximately 4\% of its total stellar mass. The origin of the gas necessary for this level of star formation remains unknown. Our data reach surface brightness limits of 29.8, 29.4, and 28.9 mag arcsec$^{-2}$ in the Sloan g, r, and i filters, respectively, and reveal no evidence of tidal features or merger remnants down to $\sim$30 mag arcsec$^{-2}$ (or equivalently, at a radial distance larger than 1.6 kpc). The galaxy's undisturbed outer elliptical morphology suggests that its recent central star formation was likely produced by either in-situ residual gas or by the accretion of external pristine gas not associated with a minor merging activity.

\end{abstract}

\keywords{galaxies: individual (NGC6789) -- galaxies: structure --- galaxies: dwarf -- galaxies: formation -- galaxies: photometry }

\section{Introduction} 

The dwarf galaxy NGC 6789 has attracted considerable attention because it shows recent central star formation despite its extreme isolation \citep{2000A&AS..142..347D}. Approximately 4\% of its total stellar mass — about 10$^8$ M$_\odot$ — formed within the past 600 Myr \citep{2010ApJ...724...49M}. This recent activity gives the galaxy’s core an irregular appearance. In fact, based on shallow imaging, it was initially classified as an irregular system (type Im) \citep{1991rc3..book.....D}. However, deeper imaging later revealed that the central star-forming region is embedded within an apparently undisturbed, redder elliptical outer structure. This finding led \citet{2000A&AS..142..347D} to propose a revised iE classification, typical of blue compact dwarf (BCD) galaxies.

A detailed spectroscopic study by \citet{2012MNRAS.423..406G} suggested that the infall of pristine, metal-poor gas could explain both the enhanced N/O ratio and the nearly simultaneous star formation episodes in multiple central regions. This raises a key question: given the galaxy’s isolation and its apparently undisturbed external shape, what is the source of the gas sustaining its recent star formation?

In this Research Note, we present substantially deeper multiband imaging of NGC 6789 to explore its outer regions and search for faint features that may reveal evidence of past minor mergers or gas accretion events capable of supplying the fuel required to build its star-forming core.

\section{TTT3 very deep imaging}

The deep imaging of NGC6789 presented in this Research Note was obtained during the commissioning of the TTT3, a 2.0-m $f$/6 Ritchey-Chrétien  telescope located at Teide Observatory. The images were obtained with COLORS, a 2k$\times$2k camera mounted at the Nasmyth 2 focus, equipped with a back-illuminated 13.5~$\mu$m pixel$^{-1}$ BEX2-DD CCD sensor, resulting in a field of view of 7.85$'$$\times$7.85$'$ and a plate scale of 0.23$''$~pixel$^{-1}$. The galaxy was observed in the Sloan filters g (2.6h), r (3.1h) and i (3.0h), reaching a limiting surface brightness (3$\sigma$; in areas equivalent to 10$^{\prime\prime}$ $\times$ 10$^{\prime\prime}$) of 29.8 mag/arcsec$^2$ (g-band),  29.4 mag/arcsec$^2$ (r-band) and  28.9 mag/arcsec$^2$ (i-band) respectively.

\begin{figure*}[ht!]
\includegraphics[width=\textwidth]{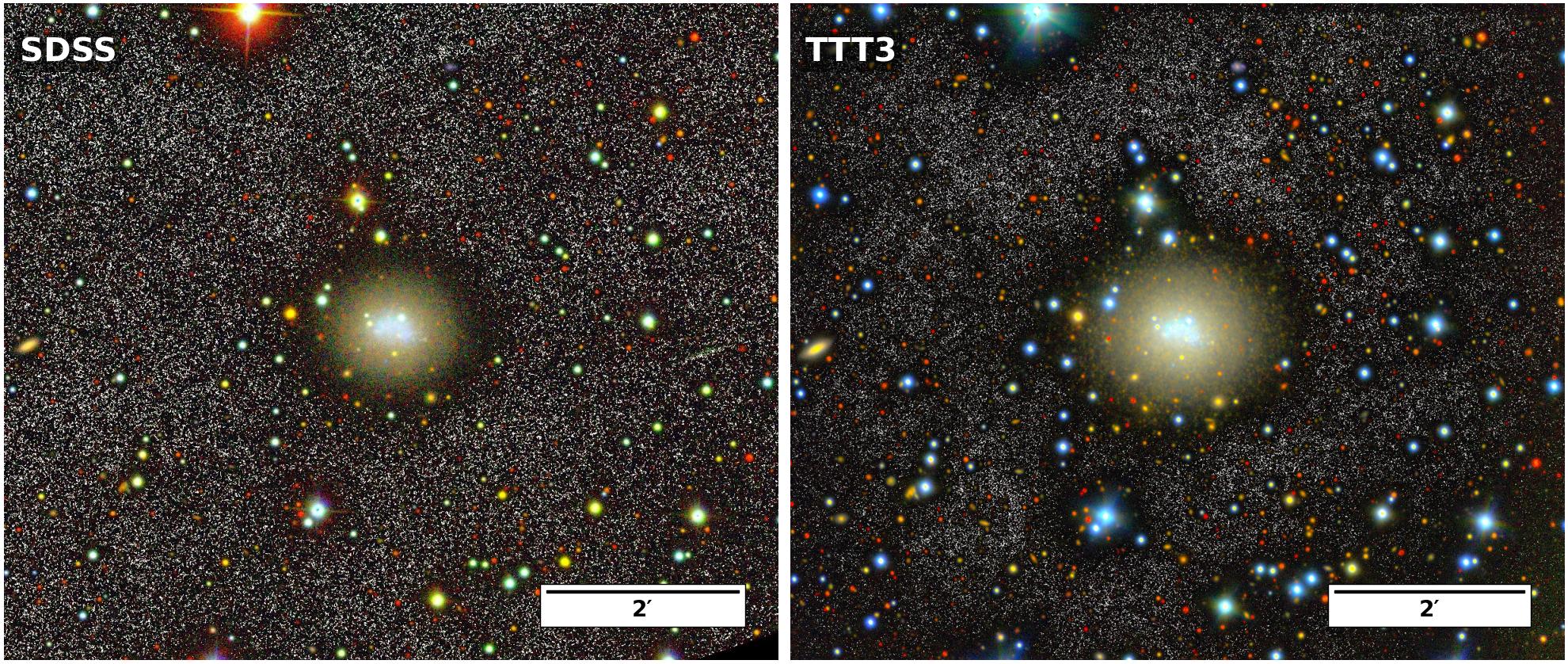}
\caption{NGC6789 as seen by the Sloan Digital Sky Survey (left panel) and the deep image obtained using  the TTT3 telescope (right panel). In both cases, the color images were created using a combination of Sloan g, r, and i filters. At the galaxy's distance of 3.6 Mpc, one arcmin corresponds to approximately 1.1 kpc. The black-and-white background of the image corresponds to the g-band image, which is our deepest dataset.
\label{fig:image}}
\end{figure*}

Figure  \ref{fig:image} shows a comparison of the usual view of the image (as seen by the SDSS) and the new deep image obtained with the TTT3. Despite the increase in depth by nearly one order of magnitude, the galaxy shows no signs of recent merging activity. The galaxy maintains its elliptical shape down to $\sim$30 mag/arcsec$^2$ (g-band) or, equivalently, a radial distance of 1.5 arcmin \citep[i.e. 1.6 kpc at 3.6 Mpc; ][]{2001ApJ...551L.135D}. The galaxy exhibits a significant color gradient due to its central star-forming region, ranging from (g-r)$_0$$\sim$0.25 in the innermost 0.25 arcmin (270 pc) to (g-r)$_0$$\sim$0.48 in the outer regions.

The lack of any observable feature brighter than 29.8 mag/arcsec$^2$ (g-band) around the galaxy places an upper limit on the amount of stellar mass that could potentially surround the galaxy as a result of the disruption of any small satellite. Assuming a stellar surface density of 0.1 M$_\odot$/pc$^2$ (based on the galaxy's outer mass density profile at around 1.5 arcmin) and an area of $\sim$2.14 kpc$^2$, corresponding to an elliptical annulus between 1.6 and 1.8 kpc, the upper limit of the stellar mass around the galaxy is $\sim$2$\times$10$^5$ M$_\odot$. Considering that the amount of new stars in the central region of the galaxy is around 4$\times$10$^6$ M$_\odot$ (and is therefore likely associated with an original amount of gas of at least 10$^7$ M$_\odot$), the absence of any visible stellar streams around the galaxy reinforces the hypothesis that the central starburst of NGC 6789 is either internally originated due to the presence of residual gas or is produced by the recent infall of pristine gas not associated with a significant stellar population.

\begin{acknowledgments}

This article is based on observations made in the Two-meter Twin Telescope (TTT3\footnote{\url{http://ttt.iac.es}}) sited at the Teide Observatory of the Instituto de Astrofísica de Canarias (IAC), that Light Bridges operates in Tenerife, Canary Islands (Spain). The observation time rights (DTO) used for this research were consumed in the PEI "GALAXDIF25". This research used storage and computing capacity in ASTRO POC's EDGE computing center at Tenerife under the form of Indefeasible Computer Rights (ICR). The ICR were consumed in the PEI “GALAXDIF25" with the collaboration of Betchle AG. Dr. Antonio Maudes’s insights in economics and law were instrumental in shaping the development of this work. Co-funded by the State Research Agency (AEI-MCINN) of the Spanish Ministry of Science and Innovation under the grant PID2022-140869NB-I00. This research also acknowledge support from the European Union through the following grants: "UNDARK" and "Excellence in Galaxies - Twinning the IAC" of the EU Horizon Europe Widening Actions programs (project numbers 101159929 and 101158446). Funding for this work/research was provided by the European Union (MSCA EDUCADO, GA 101119830). 
\end{acknowledgments}

\begin{contribution}

IT wrote the Research Note and led the interpretation of the data. SGA and IRC processed and generated the images shown. MM contributed to interpretation of the data. MRA and MSR are responsible for commissioning the TTT.

\end{contribution}

\facilities{Two-meter Twin Telescope (TTT)}

\software{Astropy \citep{Astropy},  
          Source Extractor \citep{BertinArnouts1996},
          GNuastro  \citep{Akhlaghi2019},     
          Photutils \citep{photutils},
          Numpy \citep{harris2020array},
          SWarp \citep{Bertin2010}
          SCAMP \citep{Bertin2006}
         Astrometry.net \citep{Lang2010}}


%% For this sample we use BibTeX plus aasjournalv7.bst to generate the

\bibliography{sample701.bib}{}
%\bibliographystyle{aasjournalv7}

%\bibliography{references.bib}{}
\bibliographystyle{aasjournal}



\end{document}